\documentclass[amsfonts,prb,aps]{revtex4}
\usepackage{amsmath}\usepackage{amssymb}
\begin{document}
\title{Single-order-parameter description of glass-forming liquids: A one-frequency test} 
\author{Niels L. Ellegaard, Tage Christensen, Peder Voetmann Christiansen, Niels Boye Olsen, Ulf R. Pedersen, Thomas B. Schr{\o}der, and Jeppe C. Dyre}
\affiliation{DNRF Centre ``Glass and Time,'' IMFUFA (Building 27), Department of Sciences, Roskilde University, Postbox 260, DK-4000 Roskilde,
Denmark} 
\date{\today}

\begin{abstract}
Thermo-viscoelastic linear-response functions are calculated from the master equation describing viscous liquid inherent dynamics. From the imaginary parts of the frequency-dependent isobaric specific heat, isothermal compressibility, and isobaric thermal expansion coefficient, we define a ``linear dynamic Prigogine-Defay ratio''  $\Lambda_{Tp}(\omega)$ with the property that if $\Lambda_{Tp}(\omega)=1$ at one frequency, then $\Lambda_{Tp}(\omega)$ is unity at all frequencies. This happens if and only if there is a single-order-parameter description of the thermo-viscoelastic linear responses via an order parameter (which may be non-exponential in time). Generalizations to other cases of thermodynamic control parameters than temperature and pressure are treated in an Appendix.
\end{abstract}

\newcommand{\gdag}{\mathfrak{G}}
\newcommand{\ndots}{...}
\newcommand{\ntub}[2]{#1_1,#1_2\ndots,#1_#2}
\newcommand{\df}[2]{{\frac{d#2}{d#1}}}
\newcommand{\pf}[2]{{\frac{\partial#2}{\partial#1}}}
\newcommand{\pftsp}[2]{\tsp{\pf{#1}{#2}}}
\newcommand{\pfr}[2]{\left .{\frac{\partial#2}{\partial#1}}\right|_0}
\newcommand{\pfrto}[2]{\left
.{\frac{\partial^2#2}{\partial^2#1}}\right|_0}
\newcommand{\pfrdef}{\Big|_0}
\newcommand{\pt}[3]{\left(\frac{\partial#2}{\partial#1}\right)_{#3}}
\newcommand{\ptto}[3]{\left(\frac{\partial^2#2}{\partial#1^2}\right)_{#3}}
\newcommand{\pfto}[2]{{\frac{\partial^2#2}{\partial#1^2}}}
\newcommand{\po}[1]{\frac{\partial}{\partial#1}}
\newcommand{\pfij}[3]{\frac{\partial^2#3}{\partial#1 ~\partial#2}}
\newcommand{\pfrij}[3]{\left .\frac{\partial^2#3} {\partial#1 ~\partial#2}\right|_0} 
\newcommand{\pref}[1]{Eq. (\ref{#1})}
\newcommand{\vektor}[1]{\left(\begin{array}{c}#1\end{array}\right)}
\newcommand{\nmatrix}[2]{\left(\begin{array}{#1}#2\end{array}\right)}
\newcommand{\vek}[1]{{\bf #1}} 
\newcommand{\mat}[1]{{\bf #1}}
\newcommand{\dd}[1]{\delta {#1}} 
\newcommand{\dvek}[1]{{\dd{\vek #1}}}
\newcommand{\ptc}[4] {\left(\frac{\dd #2}{\dd #1}\right)_{\dd #3\dd  #4}} 
\renewcommand{\dag}{,} 
\newcommand{\lr}[1]{\left({#1}\right)}
\newcommand{\midl}[1]{{\left<#1\right>}} 
\newcommand{\half}{\frac 1 2}
\newcommand{\ratio}{\Lambda} 
\newcommand{\mgamma}{\Gamma^{0}}
\newcommand{\mprob}{P^{0}} 
\newcommand{\improb}{\lr{P^{0}}^{-1}}
\newcommand{\mprobhalf}{\lr{P^{0}}^{\frac 1 2}}
\newcommand{\improbhalf}{\lr{P^{0}}^{-\frac 1 2}} 
\newcommand{\press}{p}
\newcommand{\jj}{{j}} 
\newcommand{\kk}{{k}} 
\newcommand{\nn}{{n}}
\newcommand{\mm}{{m}} 
\newcommand{\na}{{\alpha}} 
\newcommand{\nb}{{\beta}}
\newcommand{\nc}{{\gamma}} 
\newcommand{\nd}{{\delta}}
\newcommand{\dq}{\delta^Q} 
\newcommand{\dv}{\delta^X}
\newcommand{\linf}{L^{\infty}} 
\newcommand{\ainv}{a^{-1}}
\newcommand{\mk}[1]{#1^*} 
\newcommand{\dmk}[1]{\delta {\tilde{#1}}^*}
\newcommand{\inv}{\mathfrak{I}}


\maketitle

\section{Introduction}

A famous result of classical glass science is that if a glass-forming liquid is described by a single order parameter, the Prigogine-Defay ratio is unity. Recall that, if $\Delta c_p$ is the difference between liquid and glass isobaric specific heat per unit volume at the glass transition temperature $T_g$, $\Delta\kappa_T$ the similar liquid-glass difference of isothermal compressibilities, and $\Delta\alpha_p$ the difference of isobaric thermal expansion coefficients, the Prigogine-Defay ratio $\Pi$ is defined \cite{dav52,dav53,pri54} by

\begin{equation}\label{pdf}
  \Pi \ =\  \frac{\Delta
  c_p\Delta\kappa_T}{T_g\left(\Delta\alpha_p\right)^2}\,. 
\end{equation}
Both experimentally and theoretically the following inequality was established early on: $\Pi\geq 1$.\cite{dav52,dav53,pri54,gol63,gol64,ang76,bra85,nem95,don01}  The vast majority of reported $\Pi$'s are significantly larger than unity \cite{moy76}, and for decades the consensus has been that -- possibly with a few exceptions \cite{oel77,zol82} -- a single order parameter is not sufficient for describing glass-forming liquids.

The above picture came about as follows. The glass transition is a freezing of configurational degrees of freedom. If kinetic aspects are ignored, the glass transition has the appearance of a second order phase transition in the Ehrenfest sense with continuity of volume and entropy, but discontinuity of their thermodynamic derivatives. If the ``order parameters'' (numbers characterizing the structure) are denoted by $z_1,...,z_k$, these are frozen in the glass phase. In the equilibrium liquid phase the order parameters depend on pressure and temperature, a dependence that is determined by minimizing Gibbs free energy $G(T,p,z_1,...z_k)$:
$\partial G/\partial z_i=0$. \cite{pri54} In the classical papers by Davies and Jones from the 1950's \cite{dav52,dav53} it was shown that within this framework one always has $\Pi\geq 1$ and that $\Pi=1$ if there is just a single order parameter. The simplest dynamics for the approach to equilibrium for the order parameters, $\dot z_i=-\Lambda_i\partial G/\partial z_i$, imply exponential decays for each order parameter for small perturbations from equilibrium. Under this assumption, in the case of just one order parameter any quantity relaxes exponentially to equilibrium after a small disturbance. Exponential relaxations of the macroscopic variables are seldom observed, although there is evidence that relaxation is often intrinsically exponential so that the non-exponential behavior comes from a distribution of exponential relaxations. \cite{wen00,boh01,ric06} This reflects dynamic heterogeneity, but historically the fact that macroscopic relaxations are non-exponential was often taken as confirming the conventional wisdom that one order parameter is rarely not enough. -- Besides the reported Prigogine-Defay ratios being almost always significantly larger than unity, a further classical argument for the necessity of more than one order parameter is the following:\cite{roe77} If structure were characterized by a single order parameter, glasses with the same index of diffraction would also have all other physical properties identical, which is not the case. This does not rule out the possibility that a single order parameter is sufficient to describe the linear thermo-viscoelastic response of the viscous liquid phase, however, and it is this possibility we shall inquire into in the present paper.

Because kinetic aspects cannot be ignored, the glass transition is not a phase transition. Thus there is no exact $T_g$ and strictly speaking the $\Pi$ of \pref{pdf} is not well defined. This is because the properties of a glass depends to some extent on the cooling history; moreover, extrapolations from the glass phase are somewhat ambiguous because glass properties change slightly with time. Thus the changes in specific heat, etc., found by extrapolating glass and liquid specific heats from below and above,  respectively, to $T_g$ \cite{hod94} are not rigorously well defined. This weakens the generally accepted conclusion that in the vast majority of cases there must be more than one order parameter. 

The problem of making the Prigogine-Defay ratio rigorously well defined is solved by referring exclusively to linear-response experiments on the equilibrium viscous liquid phase.\cite{roe77,moy78,moy81} In viscous liquids thermodynamic coefficients are generally frequency dependent, $c_p=c_p(\omega)$ etc. The high-frequency limits reflect glassy behavior where relaxations do not have enough time to take place, corresponding to ``frozen'' structure. Thus for the equilibrium viscous liquid phase at any given temperature $T$ one redefines \pref{pdf} to be the ``linear Prigogine-Defay ratio'' given \cite{roe77,moy78,moy81} by

\begin{equation}\label{prigogine}
  \Pi \ =\  \frac{    \big[ c_p(\omega\rightarrow 0)-  c_p(\omega\rightarrow \infty) \big]\big[  \kappa_T(\omega\rightarrow 0)- 
  \kappa_T(\omega\rightarrow \infty)\big]}     {T\big[\alpha_p(\omega\rightarrow 0) -\alpha_p(\omega\rightarrow \infty)\big]^2}\,. 
\end{equation}

Descriptions of viscous liquids and the glass transition by means of one or more order parameter were actively discussed in the 1970's.\cite{roe77,moy78,moy81,dim74, gol75,les80} Since the $\Pi$ of \pref{pdf} is not well defined and since there are no measurements of the linear Prigogine-Defay ratio of \pref{prigogine}, the question of one or more order parameters remains open.\cite{nie97,jav06} In our view, there are a number of reasons to take seriously the possibility that some glass-forming liquids may be described by a single order parameter: 1) In early computer simulations Clarke reported Prigogine-Defay ratios close to one for a 216 particle Lennard-Jones model of Argon.\cite{cla79} This is not a highly viscous liquid, but recent extensive simulations by Sciortino and co-workers of viscous liquids confirmed this scenario by showing that structural relaxations for small temperature changes are basically controlled by a single parameter (e.g., the volume).\cite{nav02,she03,mos04,sci05} 2) Experiments monitoring loss peak frequency and loss maximum of the Johari-Goldstein dielectric beta relaxation process upon temperature jumps revealed a striking correlation between these two quantities, a correlation which is difficult to explain unless structure is parameterized by a single order parameter.\cite{dyr03} In these experiments the relaxation of beta process properties is controlled by the alpha relaxation time. This is to be expected because the alpha time is the structural relaxation time, but it is highly non-trivial that alpha and beta relaxations correlate in equilibrium viscous liquids, as shown recently by B{\"o}hmer and co-workers;\cite{boh06} this may be taken as a further indication that a single order parameter controls the behavior. 3) Experiments studying dielectric relaxation under varying temperature and pressure conditions showed that the shape of the alpha loss peak as quantified by the stretching exponent $\beta$ depends only on the loss peak frequency.\cite{rol05} This could be a consequence of there being just one order parameter, determining both alpha relaxation time and stretching. 4) Finally, we would like to mention theoretical works suggesting that the Prigogine-Defay ratio may be unity in some cases.\cite{nie97,spe99,wu99} -- More speculative, in our opinion it would not be too surprising if there is just one order parameter for liquids where time-temperature superposition applies accurately, because in these liquids linear relaxations appear to be particularly simple with a generic $\omega^{-1/2}$ high-frequency decay of, e.g., the dielectric and mechanical alpha loss peaks.\cite{ols01,jak05} 

The wide-frequency measurements required to determine $\Pi$ of \pref{prigogine} will be very difficult to perform in the foreseeable future. There are, in fact, no measurements of all three required frequency-dependent thermo-viscoelastic linear-response functions. Work at our laboratory indicates that such measurements are possible,\cite{chr06} but only over a limited frequency range. This motivates a search for an alternative to the linear Prigogine-Defay ratio. In this paper we introduce a ``linear dynamic Prigogine-Defay ratio'' (where the double prime indicates the imaginary part of the corresponding linear-response function): 

\begin{equation}\label{tp}   \ratio_{Tp}(\omega)\ =\ \frac{c_p''(\omega)\kappa_T''(\omega)}
  {T_0(\alpha_p''(\omega))^{2}}\,.
\end{equation} 

Based on the description of viscous liquid dynamics in terms of Markovian inherent dynamics in Secs. II and III we prove that $ \ratio_{Tp}(\omega)\ = 1$ if and only if the linear Prigogine-Defay ratio (\pref{prigogine}) is unity. This happens if and only if there is a single, generally non-exponential order parameter. Moreover, it is proved that if $\ratio_{Tp}(\omega)\ = 1$ at one frequency, this is true at all frequencies. In section IV we investigate the matter in a thermodynamic approach and prove that in this framework the existence of a single order parameter implies that $\ratio_{Tp}(\omega)\ = 1$ at all frequencies. In the Appendix it is shown that $\ratio_{Tp}(\omega)\ $ is just one out of a family of 4 linear dynamic Prigogine-Defay ratios larger than or equal to one, which equal unity if and only if there is a single (generally non-exponential) order parameter.

\section{Response matrix for Markovian inherent dynamics}

Adopting the energy landscape approach to viscous liquid dynamics \cite{gol69} we model the liquid (henceforth: ``system'') by the inherent dynamics consisting of jumps between the potential energy minima.\cite{pal82,sti83,sch00,sci05} This description is believed to be realistic in the highly viscous phase where there is a clear separation between the vibrational time scales (in the picosecond range) and the (alpha) relaxation time scale -- the time scale of interest here. If the system has $N$ potential energy minima, an ensemble of similar systems is described by a vector of probabilities $\vek P = (\ntub P N)$ where $P_n$ denotes the fraction of systems vibrating around energy minimum $n$. If $G_n(T,p)$ denotes the vibrational Gibbs free energy of minimum $n$, the $\vek P$-dependent Gibbs free energy is defined \cite{pal82} by

\begin{equation}\label{gibbs}
  G(T,p,\vek P)  =  \sum_n P_n   \Big( G_n(T,p)  + k_B T \ln P_n\Big)\,.
\end{equation}
Via the standard thermodynamic identities the $\vek P$-dependent entropy $S$ and volume $V$ are given by

\begin{eqnarray}\label{sogv}
   S(T,p,\vek P) &\ =\ & -\sum_{n} P_n \lr{\pf T {G_n} + k_B  \ln
   P_n}\nonumber\\ 
   V(T,p,\vek P) &\ =\ & \sum_{n} P_n \pf p {G_n}\,.
\end{eqnarray}
It is convenient to introduce the notation $\vek X=(T, -p)$ for the controlled (``input'') variables and $\vek Q= (S, V)$ for the induced responses (``output''). Thus

\begin{equation}\label{defq}
   Q_\na \ =\  -\pf{X_\na}{G},~~~\na = 1,2\,.
\end{equation}
The equilibrium probability distribution is found by minimizing $G(T,p,\vek P)$ under the constraint $\sum_n P_n=1$; this leads \cite{footnote1} to the well-known canonical probabilities (where $G(T,p)$ is the Gibbs free energy of \pref{gibbs} evaluated at the equilibrium probabilities):

\begin{equation}\label{peq}
P_n^{eq}(T,p)\,=\,
\exp\left(-\frac{G_n(T,p)-G(T,p)}{k_BT}\right)\,.    
\end{equation}

Inherent dynamics consist of transitions between minima.\cite{sch00} The inherent dynamics are modelled as a Markov process; thus the probabilities obey a standard master equation \cite{kam81}

\begin{equation}\label{master}
\dot P_n\  =\ \sum_m W_{nm} P_m  \,.
\end{equation}
The rate matrix $W$ and the equilibrium distribution $P^{eq}_n$ both depend on $T$ and $p$, and for all $n$ one has

\begin{equation}\label{eqcond}
\sum_m W_{nm}(T,p)P^{eq}_ m(T,p)\,=\,0\,.
\end{equation}

In the following the variables $T$ and $p$ are assumed to vary slightly from their average values $T_0$ and $p_0$ as arbitrary externally-controlled functions of time, resulting in small changes of $S$ and $V$. Examining perturbations around the reference state $\vek X_0 = (T_0,-p_0)$ it is convenient to introduce the notation $W_{nm}^0\equiv W_{nm}(T_0,p_0)$ and $P_{ n}^{0}\equiv P_{ n}^{eq}(T_0,p_0)$. If $\delta$ denotes perturbations from the reference state, $P_n(t)=P_n^0+\delta P_n(t)$ etc., \pref{master} implies to first order

\begin{equation}\label{masterdiff}
\delta \dot P_n \ =\ \sum_m \left(W_{nm}^0 \delta P_m + \delta W_{nm}
P_m^{0}\right)\,. 
\end{equation} 
Since $\delta( \sum_m W_{nm} P_m^{eq})=0$ (\pref{eqcond}), one has $\sum_m \delta W_{nm} P_m^0 =-\sum_m W_{nm}^0 \delta P_m^{eq}$. When this is substituted into Eq. (\ref{masterdiff}) we get

\begin{equation} \label{wdp}
\delta \dot P_n \ =\ \sum_m W_{n m}^0 \lr {\delta P_{ m}-\delta P_{m}^{eq}}\,. 
\end{equation} 
Next, we expand $\delta P_{ m}^{eq}$ in terms of $\delta X_\na$. Equations (\ref{sogv}) and (\ref{peq}) imply \cite{footnote2} (where $\beta=1,2$ and $\delta_{\nb 1}$ is the Kronecker delta symbol)

\begin{equation}\label{9}
\pf{X_\nb}{\ln   P_{ m}^{eq}}\ =\  \frac {1}{k_BT}\lr{\pf {P_m} {Q_{\nb}}
-Q_{\nb} + \delta_{\nb 1} k_B}\,. 
\end{equation} 
Thus to lowest order

\begin{equation}\label{extra}
\delta  P_m^{eq}\,=\,
\frac{P_m^{0}}{k_BT_0}\left[\left(\frac{\partial Q_1}{\partial
P_m}-Q_1+k_B\right)\delta X_1 +\left(\frac{\partial Q_2}{\partial
P_m}-Q_2\right)\delta X_2\right]\,. 
\end{equation} 
Inserting this into \pref{wdp} and utilizing \pref{eqcond} we arrive at the following ``equation of motion'' for the probabilities when temperature and pressure vary slightly as arbitrary functions of time ($\delta X_\nb\equiv\delta X_\nb(t)$):

\begin{equation}\label{dotp}
\delta \dot  P_n \ =\  \sum_m W_{n m}^{0} \left(\delta  P_ m- \frac
{1}{k_BT_0}\sum_\nb P^0_{ m}\pf  {P_m} {Q_{\nb}}\delta X_\nb \right)\,.
\end{equation}

Consider now a harmonic linear perturbation of the system. In relaxing systems like glass-forming liquids the ordinary thermodynamic response functions are frequency dependent and complex. Writing $T(t)=T_0 +{\rm Re}(\dd T~e^{i  \omega t})$ and $p(t)=p_0 + {\rm Re}(\dd p~e^{i \omega t})$, the oscillations of $S$ and $V$ are similarly described by $S(t)=S_0 +{\rm Re}(\dd S~e^{i \omega t})$ and $V(t)=V_0 +{\rm Re}(\dd V~e^{i \omega t})$. If $ \dd P_n$ and $\dd X_\nb$ denote the (complex) frequency-dependent amplitudes, in steady state Eq. (\ref{dotp}) implies that

\begin{equation}\label{laplace}
\sum_n  \lr{ W_{ln}^0-i\omega\delta_{ln}} \dd P_n  \ =\  \frac {1}{k_BT_0}\sum_{m,\nb}  W_{l m}^0  P_ m^0\pf {P_m} {Q_{\nb}}\dd X_\nb\,.
\end{equation} 
For $\omega >0$ we solve \pref{laplace} by defining (where $I$ is the identity matrix)

\begin{equation}\label{akrav}
A_{n m}(\omega) \ =\  \frac {1} {k_BT_0}   \sum_l\lr{ W^0 -i\omega I
}^{-1}_{nl}  W_{l m}^0 P_m^0 \,, 
\end{equation} 
leading to the following equation for the frequency-dependent complex probability amplitudes

\begin{equation}\label{a}
\dd P_n  \  =\  \sum_{m,\nb} A_{n m}(\omega) \pf {P_m} {Q_{\nb}} \dd
X_\nb\,. 
\end{equation} 
In order to calculate the thermo-viscoelastic linear-response functions we first define $J_{\na\nb}^{\infty}= \pf{X_\nb}{Q_\na}$. This 2$\times$2 matrix gives the instantaneous (glassy) linear response because it determines the variations of $S$ and $V$ when the probabilities $P_n$ are frozen.\cite{pal82} Next, we expand the frequency-dependent complex amplitudes $\dd Q_\na$ to get

\begin{equation}\label{ddq}  
\dd Q_\na \ =\  \sum_{n}\pf {P_n} {Q_{\na}} (T_0,p_0) \dd P_n  +
\sum_{\nb}J_{\na\nb}^{\infty}\dd X_\nb\,. 
\end{equation} 
Here and henceforth, for any function $f$ the notation $f(T_0,p_0)$ for a variable that also depends on the probabilities $\vek P$ signifies that $f$ is evaluated at the equilibrium probabilities (Eq. (\ref{peq})). When Eq. (\ref{a}) is inserted into Eq. (\ref{ddq}), we get 

\begin{equation}\label{extraeq}
\dd Q_\na\,=\,
\sum_{\nb}J_{\na\nb}(\omega)\dd X_\nb\,,
\end{equation}
where

\begin{equation}\label{dq}
J_{\na\nb}(\omega)  \ =\ J_{\na\nb}^{\infty}   +\sum_{m,n}\pf {P_n}
{Q_{\na}} (T_0,p_0)  A_{n m}(\omega)  \pf {P_m} {Q_{\nb}}(T_0,p_0) \,.  
\end{equation} 
Since $A_{n m}(\omega\rightarrow\infty) =0$ it follows that $J_{\na\nb}(\omega\rightarrow\infty)=J_{\na\nb}^\infty$, justifying the notation. 

We proceed to show that the $2\times 2$-matrix $J_{\na\nb}(\omega)$ is symmetric. Introducing the matrix $Y_{nm}$ defined by 

\begin{equation}\label{ymatrix}
Y_{nm} = (P_{n}^0)^{-\half} W_{nm}^0 (P_{m}^0)^\half\,,
\end{equation}
the detailed balance requirement implies that $Y$ is symmetric.\cite{kam81,rei98} For later use we note that $Y$ is negative semidefinite:\cite{kam81} For all vectors $x$ one has $\langle x|Y|x\rangle\leq 0$. Moreover, equality applies if and only if $x\propto (P^0)^\half$; the latter property of $Y$ expresses ``ergodicity,'' i.e., the assumption about the master equation that all states are connected by some sequence of transitions. We define a diagonal matrix $R$ by

\begin{equation}\label{qmatrix}
R_{nm} = (P_{n}^0)^{\half} \delta_{nm}\,
\end{equation}
and note that $W^0=RYR^{-1}$. This implies that

\begin{equation}\label{gamma}
A(\omega) \,=\,
\frac {1} {k_BT_0}(W^0-i\omega I)^{-1}W^0R^2\,=\,
\frac {1} {k_BT_0}\left(R(Y-i\omega I)R^{-1}\right)^{-1}RYR\,=\,
\frac {1} {k_BT_0}R(Y-i\omega I)^{-1}YR\,.
\end{equation}
Since $Y$ and $R$ are both symmetric, it follows that $A(\omega)$ is symmetric. Thus by \pref{dq} the 2$\times$2 matrix $J_{\na\nb}(\omega)$ is symmetric. 

The $J_{\na\nb}(\omega)$ matrix determines the frequency-dependent thermo-viscoelastic linear response defined as follows: If $c_p(\omega)$ denotes the isobaric heat capacity per volume, $\alpha_p(\omega)$ the isobaric thermal-expansion coefficient, and $ \kappa_T(\omega)$ the isothermal bulk compressibility, the complex frequency-dependent coefficients $\dd T$, $\dd p$, $\dd S$, and $\dd V$ are related by the following matrix (the symmetry of which reflects the symmetry of $J_{\na\nb}(\omega)$):

\begin{eqnarray}\label{respfunc}
    \vektor{\dd S\\\dd V}
    &=&V_0\nmatrix{cc}{
       c_p(\omega)/T_0 & \alpha_p(\omega)\\
       \alpha_p(\omega) & \kappa_T(\omega)
    }\vektor{\dd T\\ -\dd p }\label{thermoelastic}\,.
\end{eqnarray}
Whenever there is time-reversal invariance, the (dc, i.e. zero-frequency) symmetry implied by a Maxwell relation translates into (ac) Onsager reciprocity, a well-known result.\cite{kam81,rei98}

\section{Linear dynamic Prigogine-Defay ratio}

We proceed to calculate the linear Prigogine-Defay ratio of \pref{prigogine} from the expression for the frequency-dependent linear response functions. Since $A_{n m}(\infty)=0$, we find by inserting \pref{dq} into \pref{prigogine} (with $T=T_0$ and where $\partial S/\partial P_n\equiv\partial S/\partial P_n(T_0,p_0)$ and $\partial V/\partial P_n\equiv\partial V/\partial P_n(T_0,p_0)$ but ``$(T_0,p_0)$'' is left out for brevity)

\begin{equation}\label{prigogineto}
   \Pi \ =\  
\frac{\lr{\sum_{m,n}\pf{P_n}{S} A_{n m}(0)\pf{P_m}{S}
}\lr{\sum_{m,n}\pf{P_{n}}{V} A_{n m}(0) \pf{P_{m}}{V}}} {
\lr{\sum_{m,n}\pf{P_{n}}{S}A_{n m}(0) \pf{P_{m}}{V}}^2} \,. 
\end{equation}
Next we reason in a way analogous to that of Davies and Jones \cite{dav53} based on the Cauchy-Schwartz inequality. This inequality is the well-known mathematical theorem that if a real symmetric matrix $\vek B$ is positive or negative semidefinite, the following applies: For any vectors $x$ and $y$ one has $\langle x|\vek B|y\rangle^2\,\leq\, \langle x|\vek B|x\rangle \langle y|\vek B|y\rangle$, and for non-zero $x$ and $y$ equality applies if and only if a number $\gamma$ exists such that  $x-\gamma y\in N(\vek B)$ where $N(\vek B)$ is the kernel (null space) of $\vek B$. The matrix $A(0)$ is positive semidefinite and the kernel of $A(0)$ is the one-dimensional linear subspace spanned by the vector $(1,...,1)$.\cite{footnote4} This implies that $\Pi\geq 1$ and that $\Pi=1$ if and only if there are constants $\gamma$ and $c$ such that the following equations apply for all $n$ (compare Refs. \cite{dav53,gol64,gup76}):

\begin{equation}\label{parallel}
\pf{P_n}{V}(T_0,p_0) \ =\ \gamma \pf{P_n}{S}(T_0,p_0)  + c\,.
\end{equation}
For reasons given in the next section this situation is referred to as the single-order-parameter case.

As mentioned in the introduction, the linear Prigogine-Defay ratio of \pref{prigogine} is difficult to measure. Instead we propose to consider the ``linear dynamic Prigogine-Defay ratio'' defined in \pref{tp}. When \pref{dq} is inserted into this we find 

\begin{equation}\label{storbrok}
\ratio_{Tp}(\omega) \ =\  \frac{\lr{\sum_{m,n}\pf{P_n}{S} A_{n
m}''(\omega)\pf{P_m}{S} }\lr{\sum_{m,n}\pf{P_{n}}{V} A_{n m}''(\omega)
\pf{P_{m}}{V}}} { \lr{\sum_{m,n}\pf{P_{n}}{S} A_{n m}''(\omega)
\pf{P_{m}}{V}}^2} \,. 
\end{equation}
Because the matrix $A''(\omega)$ is negative semidefinite for all $\omega >0$,\cite{footnote4} the Cauchy-Schwartz inequality implies that $\ratio_{Tp}(\omega)\geq 1$. Again, equality applies if and only if a number $\gamma$ exists such that the vector with $n$'th component $\partial V/\partial P_n-\gamma\partial S/\partial P_n$ is in the kernel of the matrix $A''(\omega)$, which is the one-dimensional space spanned by the vector $(1,...,1)$.\cite{footnote4} Thus for all $\omega >0$ the equation $\ratio_{Tp}(\omega)= 1$ is mathematically equivalent to \pref{parallel}, and $\ratio_{Tp}(\omega)= 1$ is equivalent to $\Pi=1$. In particular, if $\ratio_{Tp}(\omega)$ is unity at one frequency, $\ratio_{Tp}(\omega)$ is unity for all frequencies -- and this happens precisely when $\Pi=1$.

\section{Why $\Pi=1$ corresponds to a single order parameter}

We define the order parameter $\varepsilon$ by 

\begin{equation}\label{dimarzio}
  \delta \varepsilon(t) \, =\, \sum_n\pf{P_n}{S}(T_0,p_0)\delta P_n(t)\,.  
\end{equation}
Combining \pref{dimarzio} with Eqs. (\ref{ddq}) and (\ref{parallel}) we find that in any linear experiment (thus calculating to lowest order) $S$ and $V$ are functions of $T$, $p$ and $\delta\varepsilon$ ($J_{12}^\infty=J_{21}^\infty$):

\begin{eqnarray}\label{order}
 \delta  S(t) &=&\,\,\,\,  \delta\varepsilon(t) +J_{11}^\infty \delta T(t)   -
 J_{12}^\infty  \delta p(t) \nonumber\\ 
\delta V(t)  &=&  \gamma\, \delta\varepsilon(t) + J_{21}^\infty \delta T(t) 
 -J_{22}^\infty \delta p(t)  \,. 
\end{eqnarray} 
The situation described by these equations is more general than the single-order parameter model of Prigogine and Meixner,\cite{pri54,mei59} because \pref{order} allows for an order parameter with non-exponential dynamics. The common physics, however, are that besides $T$ and $p$ one single number determines both entropy and volume. In the present approach $\varepsilon$ might well have a memory of the thermal prehistory.

By the definition we shall adopt here, the single-order-parameter situation applies whenever \pref{order} is obeyed for some variable $\delta\varepsilon$. In last section we proved that if $\ratio_{Tp}(\omega)=1$ applies at one frequency, then \pref{order} follows. We proceed to show that conversely, if \pref{order} applies for some order parameter $\varepsilon$, then $\ratio_{Tp}(\omega)=1$ for all $\omega$. To prove this, consider a situation with periodically varying temperature and pressure fields. In steady state $\delta S$, $\delta V$, and $\delta\varepsilon$ vary periodically with complex amplitudes $\delta S(\omega)$, $\delta V(\omega)$, and $\delta\varepsilon(\omega)$. According to \pref{order} these amplitudes are given by (for any $\omega$)

\begin{eqnarray}\label{periodic}
 \delta  S(\omega) &=& \,\,\,\, \delta\varepsilon(\omega) +J_{11}^\infty \delta
 T(\omega) - J_{12}^\infty  \delta p(\omega)\nonumber\\ 
\delta V(\omega)  &=&  \gamma\, \delta\varepsilon(\omega) + J_{21}^\infty
\delta T(\omega)  -J_{22}^\infty \delta p(\omega)  \,. 
\end{eqnarray} 
In the case where only temperature varies, comparing to \pref{respfunc} shows that the following two equations apply:

\begin{eqnarray}\label{dTrespfunc}
\frac{V_0}{T_0} c_p(\omega) &=& 
\left(\frac{\delta  S(\omega)}{ \delta T(\omega)}\right)_p
  =   \,\,\,\, \left(\frac{\delta\varepsilon(\omega)}{ \delta T(\omega)}\right)_p
  +J_{11}^\infty\nonumber\\ 
V_0\alpha_p(\omega) & =& 
\left(\frac{\delta  V(\omega)}{ \delta T(\omega)}\right)_p
  =  \gamma\left(\frac{\delta\varepsilon(\omega)}{ \delta
  T(\omega)}\right)_p +J_{21}^\infty\,.
\end{eqnarray}
Since the $J^\infty$'s are real numbers, we have 

\begin{equation}\label{tprigdef}
\gamma\frac{c_p''(\omega)}{T_0}\,=\,
\alpha_p''(\omega)\,.
\end{equation}
Similarly, if only pressure varies Eqs. (\ref{periodic}) and (\ref{respfunc}) imply 

\begin{eqnarray}\label{dprespfunc}
V_0\alpha_p(\omega) &=&  
-\left(\frac{\delta  S(\omega)}{ \delta p(\omega)}\right)_T
  = \,\,\,\, -\left(\frac{\delta\varepsilon(\omega)}{ \delta p(\omega)}\right)_T
  +J_{12}^\infty\nonumber\\ 
V_0\kappa_T(\omega) & =& 
-\left(\frac{\delta  V(\omega)}{ \delta p(\omega)}\right)_T
  =  -\gamma\left(\frac{\delta\varepsilon(\omega)}{ \delta
  p(\omega)}\right)_T +J_{22}^\infty\,;
\end{eqnarray}
thus 

\begin{equation}\label{pprigdef}
\gamma\alpha_p''(\omega)\,=\,
\kappa_T''(\omega)\,.
\end{equation}
Eliminating $\gamma$ by combining Eqs. (\ref{tprigdef}) and (\ref{pprigdef}) yields $ \ratio_{Tp}(\omega)=1$ for any $\omega$. Note that in the case of one order parameter the imaginary (``loss'') parts of all three response functions are proportional.

The above line of reasoning can be repeated for different choices of input and output variables. Altogether there are 4 different natural (linear) dynamic Prigogine-Defay ratios as demonstrated in the Appendix; here it is also shown that the requirement of positive dissipation implies that no dynamic Prigogine-Defay ratio can be smaller than unity.

\section{Concluding remarks}

The original Prigogine-Defay ratio of \pref{pdf} is not rigorously well defined. The ``linear'' Prigogine-Defay ratio of \pref{prigogine} {\it is} well defined, but requires measurements of thermo-viscoelastic response functions over many decades of frequency. There are still no methods for measuring a complete set of these response functions. Hopefully such measurements are possible in the near future, but realistically they will initially only cover a few decades.\cite{chr06} This motivates the one-frequency criterion proposed and developed mathematically in this paper: If the linear dynamic Prigogine-Defay ratio is unity at one frequency, this quantity is unity at all frequencies and a single-order parameter description applies. Conversely, if a single-order parameter description applies, the dynamic Prigogine-Defay ratio is unity at all frequencies. If this happens, the imaginary parts of the three linear response functions of \pref{respfunc} are proportional. As shown in the Appendix, these results all generalize to the 3 other natural choices of input and output variables. -- These are strict mathematical statements; in practice one cannot determine whether or not the dynamic Prigogine-Defay ratio is precisely one. What one can do is test how close to unity this quantity is. We expect that a single-parameter description is a good approximation whenever the dynamic Prigogine-Defay ratio is close to unity. Note that if there are several relaxation processes with well separated time scales, it is possible that some of these are well described by a single order parameter, though not perfectly, whereas others are not. In this case the frequency-dependent one-parameter test developed here in principle would be useful although, as mentioned already, it will probably not be possible in the foreseeable future to determine all three required response functions over wide frequency ranges. 

In the present paper we referred to the $ \ratio_{Tp}=1$ situation as that of a single, generally non-exponential order parameter. As emphasized by Goldstein,\cite{gol64} however, due to a mathematical equivalence it is really a matter of taste whether one prefers instead to refer to a situation of several order parameters obeying a mathematical constraint (in our case \pref{parallel}). It is not possible {\it a priori} to estimate how restrictive or unlikely it is that \pref{parallel} applies; only experiment can settle this question. It should be noted, though, that there is a simple physical interpretation of the single-order parameter case: In the approximation where each inherent state is regarded as a potential energy minimum with a harmonic potential, via Eqs. (\ref{sogv}) and (\ref{peq}) \pref{parallel} is equivalent to $V_n=\alpha_1 E_n + \alpha_2$ where $V_n$ is the inherent state volume and $E_n$ the inherent state energy (potential energy minimum). Thus there is a single order parameter if and only if inherent state volume correlates perfectly and linearly with inherent state energy. It is expected that this condition is obeyed to a good approximation if and only if the dynamic Prigogine-Defay ratio is close to unity.

\acknowledgments{The authors wish to thank Nick Bailey for helpful comments. This work was supported the Danish National Research Foundation's (DNRF) centre for viscous liquid dynamics ``Glass and Time.''}

\appendix*\section{Generalizations to other control variables}

Standard thermodynamics give rise to a number of (dc) linear-response coefficients. If the thermodynamic variables of interest are $T, p, S, V$, there are 24 coefficients of the form $(\partial a/\partial b)_c$ with $a,b,$ and $c$  chosen among $T, p, S, V$.\cite{ber78} These coefficients form 12 pairs that are trivially related by inversion [e.g., $(\partial a/\partial b)_c=1/(\partial b/\partial a) _c$]. As is well known, the 12 coefficients are not all independent, but related by various Maxwell relations (summarized below). There are the following 8 basic linear-response coefficients (where the specific heats are per unit volume and the last three coefficients have no generally accepted names): 

\begin{eqnarray}\label{thermo}
   \begin{array}{ll}
\mbox{Isochoric specific heat:} 
& c_V \equiv \frac{T}{V} \Big(\frac{\partial S}{\partial T}\Big)_V \\ 
\mbox{Isobaric  specific heat:} 
& c_p \equiv  \frac{T}{V}   \Big(\frac{\partial S}{\partial T}\Big)_p \\ 
\mbox{Isothermal  compressibility:} 
& \kappa_T \equiv -\frac{1}{V} \Big(\frac{\partial V}{\partial p}\Big)_T \\ 
\mbox{Adiabatic compressibility:} 
& \kappa_S \equiv -\frac{1}{V} \Big(\frac{\partial V}{\partial p}\Big)_S \\ 
\mbox{Isobaric expansion coefficient:} 
& \alpha_p \equiv  \frac{1}{V} \Big(\frac{\partial V}{\partial T}\Big)_p  = -\frac{1}{V}  \Big(\frac{\partial S}{\partial p}\Big)_T\\
\mbox{``Adiabatic contraction coefficient:''} 
& \alpha_S \equiv -\frac{1}{V} \Big(\frac{\partial V}{\partial T}\Big)_S = \frac{1}{V} \Big(\frac{\partial  S}{\partial p}\Big)_V\\
\mbox{``Isochoric pressure coefficient:''} 
& \beta_V \equiv \Big(\frac{\partial p}{\partial T}\Big)_V  = \Big(\frac{\partial S}{\partial V}\Big)_T \\
\mbox{``Adiabatic pressure coefficient:''} 
& \beta_S \equiv  \Big(\frac{\partial p}{\partial T}\Big)_S  = \Big(\frac{\partial S}{\partial   V}\Big)_p
\end{array}
\end{eqnarray}

Consider harmonically varying scalar thermal and mechanical perturbations of equilibrium for a small volume element. ``Small'' here means that its linear dimensions are much smaller than both the heat diffusion length and the sound wavelength corresponding to the frequency under consideration, implying that the perturbations may be regarded as homogeneous over the small volume element. Let  $\delta T(\omega),\, \delta p(\omega),\, \delta s(\omega),\, \delta v(\omega)$ denote the complex amplitudes of perturbations varying with time as $\propto \exp(i \omega t)$ where we define the intensive variables $v \equiv  V/V_0$ and $s \equiv S/V_0$ ($V_0$ is the equilibrium volume). If small perturbations around temperature $T_0$ are considered, following \pref{thermo} one defines the following complex frequency-dependent linear-response quantities (where according to the so-called correspondence principle \cite{chr82} all thermodynamic relations survive and the Maxwell relations become Onsager reciprocity relations \cite{mei59}):

\begin{eqnarray}
   \begin{array}{ll}
\mbox{Isochoric specific heat:} 
& c_V(\omega) \equiv T_0 \Big(\frac{\delta s(\omega)}{\delta T(\omega)}\Big)_V \\ 
\mbox{Isobaric  specific heat:} 
& c_p(\omega) \equiv T_0  \Big(\frac{\delta s(\omega)}{\delta T(\omega)}\Big)_p \\ 
\mbox{Isothermal  compressibility:} 
& \kappa_T (\omega)\equiv -\Big(\frac{\delta v(\omega)}{\delta p(\omega)}\Big)_T \\ 
\mbox{Adiabatic compressibility:} 
& \kappa_S (\omega)\equiv -\Big(\frac{\delta v(\omega)}{\delta p(\omega)}\Big)_S \\ 
\mbox{Isobaric expansion coefficient:} 
& \alpha_p (\omega)\equiv  \Big(\frac{\delta v(\omega)}{\delta T(\omega)}\Big)_p  = - \Big(\frac{\delta s(\omega)}{\delta p(\omega)}\Big)_T\\
\mbox{``Adiabatic contraction coefficient:''} 
& \alpha_S(\omega) \equiv -\Big(\frac{\delta v(\omega)}{\delta T(\omega)}\Big)_S = \Big(\frac{\delta  s(\omega)}{\delta p(\omega)}\Big)_V\\
\mbox{``Isochoric pressure coefficient:''} 
& \beta_V (\omega)\equiv \Big(\frac{\delta p(\omega)}{\delta T(\omega)}\Big)_V  = \Big(\frac{\delta s(\omega)}{\delta v(\omega)}\Big)_T \\
\mbox{``Adiabatic pressure coefficient:''} 
& \beta_S(\omega) \equiv  \Big(\frac{\delta p(\omega)}{\delta T(\omega)}\Big)_S  = \Big(\frac{\delta s(\omega)}{\delta v(\omega)}\Big)_p
\end{array}
\end{eqnarray}

The general procedure now works as follows. Let $(\delta X(\omega), \delta Y(\omega)$) denote the amplitudes of two periodically varying thermodynamic variables considered as control variables and $(\delta Z(\omega), \delta W(\omega)$) the remaining two periodically varying variables. The relationship between the two sets of variables generally takes the form

\begin{eqnarray}\label{generalrespfunc}
    \vektor{\dd Z(\omega)\\\dd W(\omega)}
    &=&\nmatrix{cc}{
       a_{11}(\omega) & a_{12}(\omega)\\
       a_{21}(\omega) & a_{22}(\omega)
    }\vektor{\dd X(\omega)\\ \dd Y(\omega) }\,.
\end{eqnarray}
In the case of a single order parameter $\varepsilon$ -- compare to \pref{periodic} -- we can write (where the $2\times 3$-matrix is real and frequency independent): 
\begin{eqnarray}\label{generalrespfunc1op}
    \vektor{\dd Z(\omega)\\\dd W(\omega)}
    &=&\nmatrix{ccc}{
       a^\infty_{11} & a^\infty_{12} & b_1 \\
       a^\infty_{21} & a^\infty_{22} & b_2 \\
    }\vektor{\dd X(\omega)\\ \dd Y(\omega) \\ \dd \varepsilon(\omega)}\,.
\end{eqnarray}
Comparing to \pref{generalrespfunc}, when $Y$ does not vary we get:

\begin{eqnarray}
   \begin{array}{l}\label{gamma1}
      a_{11}(\omega) =  a^\infty_{11} 
                      + b_1\left(\frac{\dd \varepsilon(\omega)}{\dd X(\omega)}\right)_Y\\
      a_{21}(\omega) =  a^\infty_{21}
                      + b_2\left(\frac{\dd \varepsilon(\omega)}{\dd X(\omega)}\right)_Y
   \end{array}
   \Big\} \Rightarrow
   \frac{a^{''}_{21}(\omega)}{a^{''}_{11}(\omega)} = \frac{b_2}{b_1}\equiv\gamma_{XY}\,,
\end{eqnarray}
and when $X$ does not vary:

\begin{eqnarray}\label{gamma2}
   \begin{array}{l}
      a_{12}(\omega) =  a^\infty_{12} 
                      + b_1\left(\frac{\dd \varepsilon(\omega)}{\dd Y(\omega)}\right)_X\\
      a_{22}(\omega) =  a^\infty_{22}
                      + b_2\left(\frac{\dd \varepsilon(\omega)}{\dd Y(\omega)}\right)_X
   \end{array}
   \Big\} \Rightarrow
   \frac{a^{''}_{22}(\omega)}{a^{''}_{12}(\omega)} = \frac{b_2}{b_1}= \gamma_{XY}\,.
\end{eqnarray}
In principle there are 6 different choices of control variables. Below we treat the 4 natural cases where the input variables consist of one from the $(S,T)$ ``energy bond'' \cite{ost73,mik93} and one from the $(-p,V)$ energy bond, and similarly for the output variables. In each case the signs are chosen to make the response matrix symmetric. Via \pref{gamma1} and \pref{gamma2} in each of the 4 cases the 3 imaginary (loss) parts of the relevant linear response functions are proportional. The 4 cases are detailed below, where the explicit amplitude frequency dependence is left out for simplicity of notation.\newline
{\bf 1. Control variables $\dd T$ and $-\dd p$ .}\newline
\vspace{0.5cm}
This case was considered in the main paper, but is included for completeness:
\begin{eqnarray}\label{respfuncTp}
    \vektor{\dd s\\\dd v}
    &=&\nmatrix{cc}{
       c_p(\omega)/T_0 & \alpha_p(\omega)\\
       \alpha_p(\omega) & \kappa_T(\omega)
    }\vektor{\dd T\\ -\dd p }\,.
\end{eqnarray}
Applying \pref{gamma1} and \pref{gamma2} to this we get
\begin{eqnarray}
  \frac{T_0\alpha^{''}_p(\omega)}{c^{''}_p(\omega)} 
   = \gamma_{Tp}
   =  \frac{\kappa^{''}_T(\omega)}{\alpha^{''}_p(\omega)},
\end{eqnarray}
and thus
\begin{eqnarray}
  \Lambda_{Tp}(\omega) \equiv 
        \frac{c^{''}_p(\omega)\kappa^{''}_T(\omega)}
             {T_0 \left[ \alpha^{''}_p(\omega)\right]^2 } = 1.
\end{eqnarray}
{\bf 2. Control variables $\dd s$ and $\dd v$ :}

\begin{eqnarray}\label{respfuncSV}
    \vektor{\dd T\\ -\dd p}
    &=&\nmatrix{cc}{
       T_0/c_V(\omega) & -1/\alpha_S(\omega)\\
       -1/\alpha_S(\omega) & 1/\kappa_S(\omega)
    }\vektor{\dd s\\ \dd v}\,.
\end{eqnarray}
Applying Eqs. (\ref{gamma1}) and (\ref{gamma2}) to this we get
\begin{eqnarray}
  -\frac{\left(1/\alpha_S(\omega)\right)^{''}}{\left(T_0/c_V(\omega)\right
  )^{''}} = \gamma_{SV} = 
  -\frac{\left(1/\kappa_S(\omega)\right)^{''}}{\left(1/\alpha_S(\omega)\right)^{''}}
\end{eqnarray}
and thus
\begin{eqnarray}
  \Lambda_{SV}(\omega) \equiv 
        \frac{\left(T_0/c_V(\omega)\right)^{''}
        \left(1/\kappa_S(\omega)\right)^{''}}
             { \left[ \left(1/\alpha_S(\omega)\right)^{''} \right]^2} = 1.
\end{eqnarray}
{\bf 3. Control variables $\dd s$ and $-\dd p$ :}

\begin{eqnarray}\label{respfuncSp}
    \vektor{\dd T\\ -\dd v}
    &=&\nmatrix{cc}{
       T_0/c_p(\omega) & -1/\beta_S(\omega)\\
       -1/\beta_S(\omega) & -\kappa_S(\omega)
    }\vektor{\dd s\\ -\dd p}\,.
\end{eqnarray}
Applying \pref{gamma1} and \pref{gamma2} to this we get
\begin{eqnarray}
  -\frac{\left(1/\beta_S(\omega)\right)^{''}}{\left(T_0/c_p(\omega)\right)
  ^{''}} = \gamma_{Sp} = \frac{\kappa^{''}_S(\omega)}
  {\left(1/\beta_S(\omega)\right)^{''}}
\end{eqnarray}
and thus
\begin{eqnarray}
  \Lambda_{Sp}(\omega) \equiv 
       - \frac{\left(T_0/c_p(\omega)\right)^{''} \kappa^{''}_S(\omega)}
             { \left[ \left(1/\beta_S(\omega)\right)^{''} \right]^2} = 1.
\end{eqnarray}
{\bf 4. Control variables $\dd T$ and $\dd v$ :}

\begin{eqnarray}\label{respfuncTV}
    \vektor{\dd s\\ -\dd p}
    &=&\nmatrix{cc}{
       c_V(\omega)/T_0 & -\beta_V(\omega)\\
       -\beta_V(\omega) & -1/\kappa_T(\omega)
    }\vektor{\dd T\\ -\dd v}\,.
\end{eqnarray}
Applying \pref{gamma1} and \pref{gamma2} to this we get
\begin{eqnarray}
  -\frac{T_0\beta^{''}_V(\omega)}{c^{''}_V(\omega)} 
  = \gamma_{TV}
  = \frac{\left(1/\kappa_T(\omega)\right)^{''}}{\beta^{''}_V(\omega)}
\end{eqnarray} and thus \begin{eqnarray}
  \Lambda_{TV}(\omega) \equiv 
        -\frac{c^{''}_V(\omega) \left(1/\kappa_T(\omega)\right)^{''}}
             { T_0 \left[ \beta^{''}_V(\omega) \right]^2} = 1.
\end{eqnarray}

For each case we showed above that, if there is a single order parameter, the dynamic Prigogine-Defay ratio is unity. Below we proceed to prove that, conversely, if one of the dynamic Prigogine-Defay ratios is unity at some frequency, then this dynamic Prigogine-Defay ratio is unity at all frequencies and there is a single order parameter. In particular, if one dynamic Prigogine-Defay ratio is unity at one frequency, all dynamic Prigogine-Defay ratios are unity at all frequencies.

In the periodic situation the dissipation is proportional to ${\rm Im} [\dd T( \omega)\dd s^*( \omega)- \dd p( \omega)\dd v^*( \omega)]$.\cite{mei59,bat79} The requirement of positive dissipation implies that all 4 dynamic Prigogine-Defay ratios are larger than or equal to unity; this follows by considering the special case of in-phase input variables in which case it is easy to show that the determinant of the imaginary response matrix must be non-negative to have positive dissipation. If one of the dynamic Prigogine-Defay ratios is unity at some frequency $\omega$, the determinant of the imaginary part of the corresponding response matrix is zero. This implies that there is an eigenvector of the imaginary response matrix with zero eigenvalue. Thus the dynamic Prigogine-Defay ratio is unity if and only if a thermodynamic cycle exists with zero dissipation. It follows that, if in one of cases 2-4 the dynamic Prigogine-Defay ratio is unity at frequency $\omega$, then $\Lambda_{Tp}(\omega)=1$. This implies that $\Lambda_{Tp}(\omega)=1$ at all frequencies (Sec. III), and that there is a single order parameter $\varepsilon$ such that $S$ and $V$ are given by \pref{order}. These equations are easily rewritten to give the required output variables in terms of the input variables and $\varepsilon$. This implies that the relevant dynamic Prigogine-Defay ratio is unity at all frequencies, and that all other dynamic Prigogine-Defay ratio are also unity at all frequencies.


\begin{thebibliography}{99}
\bibitem{footnote1} The relation $\pf {P_n} G = {\rm const.}$ leads to $G_n + k_B T\ln   P_n^{eq} + k_BT = {\rm const}$. Insert this into \pref{gibbs} to find $G_n + k_B   T\ln P_n^{eq} = G$, then isolate $P_n^{eq}$ to get the desired result.
\bibitem{footnote4} If the eigenvalues of $Y$ are denoted by $\lambda_k$ with corresponding eigenvectors $e_k$, there are $N-1$ negative eigenvalues and one zero eigenvalue.\cite{kam81}  \pref{gamma} implies that for any vector $x$ we have $\langle x|A(\omega)|x\rangle=(1/k_BT_0)\sum_{\lambda_k<0}\langle Rx|e_k\rangle\langle e_k|Rx\rangle\lambda_k/(\lambda_k-i\omega)$ (where the sum is restricted to the $N-1$ negative eigenstates). Letting $\omega\rightarrow 0$ we find that $\langle x|A(0)|x\rangle\geq 0$, showing that $A(0)$ is positive semidefinite. For the imaginary part, we find that $\langle x|A''(\omega)|x\rangle=(1/k_BT_0)\sum_{\lambda_k<0}\langle Rx|e_k\rangle\langle e_k|Rx\rangle\lambda_k\omega/(\lambda_k^2+\omega^2)\leq 0$ whenever $\omega>0$; thus $A''(\omega)$ is negative semidefinite. In both cases, equality applies if and only if the vector $Rx$ lies in the one-dimensional eigenspace of $Y$ corresponding to the zero eigenvalue. Thus equality applies if and only if $(Rx)_n\propto (P_n^0)^\half$, or $x\propto (1, ... , 1)$.
\end{thebibliography}
\end{document}